\documentclass[sigconf]{acmart}

\usepackage{booktabs} 
\usepackage{multirow}
\usepackage{balance} 
\usepackage{adjustbox}
\usepackage{footnote}

\copyrightyear{2018}
\acmYear{2018}
\setcopyright{rightsretained}
\acmConference[CAIR'18]{Second International Workshop on Conversational Approaches to Information Retrieval}{July 12, 2018}{Ann Arbor Michigan, USA}
\acmBooktitle{Second International Workshop on Conversational Approaches to Information Retrieval (CAIR'18), July 12, 2018, Ann Arbor Michigan, USA}
\acmPrice{15.00}
\acmDOI{xx.xxxx/xx.xx}
\acmISBN{xx-x-xxxx-xxxx-x/xx/xx}

\acmArticle{4}

\begin{document}

\title[Exploring Current User Web Search Behaviours]{Exploring Current User Web Search Behaviours in Analysis Tasks to be Supported in Conversational Search}


\author{Abhishek Kaushik}
\affiliation{%
  \institution{ADAPT Centre, School of Computing \\ 
  Dublin City University \\
  Dublin 9, Ireland}
}
\email{abhishek.kaushik2@mail.dcu.ie}

\author{Gareth J. F. Jones}
\affiliation{%
  \institution{ADAPT Centre, School of Computing \\ 
  Dublin City University \\
  Dublin 9, Ireland}
}

\email{Gareth.Jones@dcu.ie}

\begin{abstract}
Conversational search presents opportunities to support users in their search activities to improve the effectiveness and efficiency of search while reducing their cognitive load. Limitations of the potential competency of conversational agents restrict the situations for which conversational search agents can replace human intermediaries. It is thus more interesting, initially at least, to investigate opportunities for conversational interaction to support less complex information retrieval tasks, such as typical web search, which do not require human level intelligence in the conversational agent. In order to move towards the development of a system to enable conversational search of this type, we need to understand their required capabilities. To progress our understanding of these, we report a study examining the behaviour of users when using a standard web search engine, designed to enable us to identify opportunities to support their search activities using a conversational agent.
\end{abstract}



\maketitle

\section{Introduction}

One of the motivating objectives of {\em conversational search\/} is to support more natural interaction between users and content information archives via dialogue-based engagement. The ideal conversational interface might perhaps be seen as an automated surrogate for a human assistant with the competence to support search with the intelligence of such an assistant. An aspiration of this sort is, at least in the short to medium term, unrealistic. Search tasks requiring this sort of support will continue to need to engage human intermediaries to satisfy the searcher's information need. At this time, research in conversational search should seek to support information retrieval (IR) activities where users work, or could work, unassisted with search applications. For example, to enable more effective or efficient engagement with web search engines, reducing the cognitive load on the searcher necessary to achieve their objective, while improving search effectiveness. The entity with which the user engages would be a surrogate for a less skilled human intermediary, which can be referred as an {\em agent\/}. The question then arises, in order to provide useful support to a user's search activities, what does the conversational agent need to be able to do? In agent terms, what {\em competence\/} does it need? How does the user learn to {\em trust\/} the activities of the agent?

Support of web search requires the conversational agent be able to handle queries relating an information need in any topic area of interest to the searcher. Information needs in these areas can be of different forms, ranging for example from straightforward factoid questions to more exploratory needs requiring consultation of multiple retrieved items. A further factor is the varying knowledge of the subject area of the searcher, and their consequential interest in different items from available in the information archive.

In order to begin to specify the necessary functionality of a conversational search agent in this setting, we need to understand what users seek to do in order to satisfy their information needs, and to study their current actions as they try to achieve this. Models of information seeking, such as those of Belkin \cite{belkin1982}, Kuhlthau \cite{Kuhlthau1991InsideTS}, and Vakkari \cite{vakkari2016searching}, provide mechanisms for understanding the cognitive processes associated with satisfying a user's information need. Regardless of the interaction mechanisms they support, search applications seek to provide mechanisms to resolve information needs. The manner in which the user learns about the subject, or progresses in the resolution of their information need, will depend on the manner of their interaction with the search application. Current applications rely on the user to create their search query, and to work with retrieved surrogates such as snippets and potentially with query reformulation and further searches, to address their information needs. The argument for conversational search is that it has the potential to better direct completion of search tasks to improve efficiency, reduce cognitive load or user effort, and potentially to enable more successful completion of search tasks by providing proactive support than users are able to do with current entirely user driven tools.


In this paper we report an initial study examining user search behaviour as they seek to satisfy a predefined information need specified in a backstory using standard engagement with a commercial search engine. This study contrasts with other studies which have approached understanding the potential for conversational search by studying search by using two individuals with one taking he role of the searcher and the other the search agent intermediary. The results of our study are designed to provide input into the design of a conversational search agent to support common web search tasks.


The remainder of this paper is structured as follows: Section \ref{existing} considers agents and existing user studies relating to conversational search, Section \ref{models} briefly introduces information seeking models and learning in search,  Section \ref{methodology} describes the methodology of our study, Sections \ref{results1} and \ref{results2} present results and analysis from our study, and finally Section \ref{conclusions} concludes and outlines directions for the next stage of our work. 

%
%



\section{Agents and Users Studies of Conversational Search}
\label{existing}

In this section we consider the deployment of agents and their role in conversational search, and briefly review previous studies in user behaviour for human mediated search.

\subsection{Intelligent Agents}

Intelligent agents have been an area of study for many years, while there is a considerable literature exploring many topics in agent technology, a small number of key concepts continue to underlie their specification and utilization. Key among these are the concepts of: {\em competence\/}, what is the agent able to do, and how does it acquire the knowledge to do this?; {\em trust\/}, how can we ensure that the user feels comfortable delegating tasks to the agent? \cite{Maes:1994:ARW:176789.176792}.

When designed to assist a human with a task, as is the case with conversation search, a principal consideration is the location of the agent within the information flow, and the extent to which the agent has autonomy and to which the user has control. Placing the agent between the user and the search engine means that the agent has complete control, and the user must rely on the competence of the agent and trust it to carry out or assist in the search effectively. Placing the agent alongside the searcher gives the potential for joint control of the search process. This latter arrangement is often referred by the metaphor of the {\em personal assistant\/}. In terms of conversational search, we can think of the agent taking the role of human intermediary, such as a librarian. The agent assists the user in their search, with the communication taking the form of a conversation or dialogue. In different situations the agent make take a different position in this arrangement, with a differing level of autonomy. 

\subsection{Studies of Intermediary-Based Search}

The concept of computer-mediated dialogue in search is a longstanding goal within IR research, although until recently not one which has received significant attention. Within this work, one area of study is understanding the processes of a user engaging in search in conjunction with a human intermediary. An early key study reporting a study of this type is described in \cite{Daniels1985}. The goal was to work towards a model to enable the specification of an {\em information provision mechanism\/} which cooperates with the user in an information seeking task. This work was set in the context of academic researchers seeking information in libraries with the support of professional librarians. Much more recently work by Trippas et al at RMIT is performing similar types of studies focusing on speech only based search for users seeking information on much more general topics of the type that a user might pose to a web search engine \cite{Trippas2018}\cite{Trippas:2017:PIC:3020165.3022144}.

While understanding the conversational behaviour of searchers in fulfilling their information needs with the assistance of expert searchers, and of the potential for human-like intermediaries to assist with web search type activities is of course important, implementation of a conversational agent able to support search in this way would require a highly competent agent. In this study, we approach conversational search from the contrasting objective of understanding current web search type activities as undertaken by users without intermediary assistance with the aim of identifying opportunities for the inclusion of conversational assistance to support them.

\section{Information Seeking and Learning Models}
\label{models}

Observing the behaviour of searchers and creating a conversational agent to work as a collaborative assistant to supporting them in satisfying their information needs could be based on freestanding rule-based or statistical models. However, a more complete, extensible solution is likely to be achieved if their activities are placed within an information seeking model around which the competence of an information agent can be modeled. 

Significant early contributions to understanding and modeling the information needs and activities of searchers were made in Belkin's work in the development of the {\em Anomalous States of Knowledge (ASK)\/} model \cite{belkin1980anomalous}, which highlighted the difficulties of users in specifying the details of their information needs for engagement with IR systems. The ASK model in IR was further elaborated in \cite{belkin1982}. Various extensions to the ASK model and models taking alternative or complementary approaches to modeling information seeking have been developed since then. These models include Kuhlthau's {\em Information Search Process (ISP)\/} model which captures the searcher's constructive activity of examining retrieved information in order to extend their state of knowledge as they seek to resolve their information need \cite{Kuhlthau1991InsideTS}. In ISP, information seeking is viewed as an active process in which the user attempts to find new information and integrate it with what they already know. Essentially this is a personal activity of sense-making as the user understands and interprets the information. The process concludes when the user either satisfies their information needs or develops new understanding leading to a reformatted query or possibly abandonment of the search task. In this study we focus on Vakkari's analysis of {\em Search as Learning\/} model \cite{vakkari2016searching}. In this model search is viewed as a process incorporating learning, in which engagement with retrieved information leads to revisions in the searcher's personal knowledge structures relating to the topic under investigation. The form of the restructuring, modification or elaboration of the user's knowledge as the search progresses depends on their existing knowledge of this or related subjects. Specifically Vakkari identifies three forms of change in the searcher's knowledge structure associated with a search task: 

\begin{itemize}

\item {\bf Restructuring\/}: Occurs when the searcher's knowledge of the subject is vague and lacking structure. In this state the searcher is likely to examine multiple sources and potentially have difficulty identifying relevant information without extended investigation. 

\item {\bf Tuning\/}: During which the searcher stabilizes and extends their knowledge of the topic. They may examine less sources, but they are likely to have sufficient knowledge to select them more discriminatively and on average to examine them in more detail.

\item{\bf Assimilation\/}: Where the searcher combines new knowledge with their existing knowledge structures.

\end{itemize}

The activities of learning about a topic during exploratory search provide an obvious setting in which the searcher could progress between these phases as they learn about aspects of a topic of interest \cite{qu2008}. It is unlikely that there will be a smooth transition between these phases, since the searcher is likely to have different levels of knowledge relating to different aspects of the overall knowledge space needed to resolve their information need. Thus, as noted by studies of information seeking, this is a complex cognitive process, but one which needs to be properly considered in the design of conversational search agents. In order to being our study of how searcher address information needs with differing levels of knowledge of the topic of their search, and to enable us to explore how conversational agents might support, we conducted a study examining searchers interactions with a standard web search engine. The details of this study are described in the following sections.

\section{Experimental Methodology}
\label{methodology}

In this section we describe the details of our user study. 
The study aimed to enable us to observe and better understand the behaviour of non-specialist searchers whose techniques for use of search engines are generally learned from personal experience. The objective of this study is to provide input into specification of potential conversational extensions to existing search applications. 
We seek to gain insights not only into how conversation engagement might directly be incorporated into their current search activities, but also where there are opportunities to enhance the user's search experience.

\subsection{Information Needs for Study}

For our search tasks we wished to give searchers realistic information needs which could be satisfied using a standard web search engine. In order to control the form and detail of these, we decided to use a set of information needs specified within {\em backstories\/}. The backstories that we selected where taken from the UQV100 test collection \cite{Bailey:2016:UTC:2911451.2914671}. This consists of 100 backstories created from the TREC 2013 and 2014 Web tracks. In previous work, the creators of the UQV100 test collection developed an earlier collection of backstories in which they classified search tasks according to their level of cognitive complexity based on the Taxonomy of Learning \cite{krathwohl2002revision}, from which they adopted three of the five cognitive dimensions:
{\em Remember\/}, {\em Understand\/} and {\em Analyze\/} defined as follows: 
\begin{itemize}
\item Remember: retrieving, recalling and recognizing the information from memory.

\item Understand: developing the meaning of verbal and non-verbal communication through exemplifying, explaining, interpreting, classifying, comparing and inferring.

\item Analyze: dividing the problem into small parts, examining how each part relates to each other, and to an overall structure through breaking, organizing, and attributing.

\end{itemize}

The search tasks in this earlier study were based on TREC tasks from 2002, 2003 and 2004, and were labeled with their level of cognitive complexity \cite{bailey2015user}. 
While all three types may benefit from search using conversational interaction, we decided to focus on the most cognitively engaging {\em Analyze\/} type in the expectation that this would require a greater level of activity to satisfy the information need.
Since the UQV100 topics were not provided with these labels, we selected a suitable subset as follows. The UQV100 topics were provided labeled with estimates of the number of queries would need to be entered and the number of documents that would need to be accessed in order to satisfy them. We used the product of these figures as an estimate of the expected cognitive complexity, and then manually selected 12 of the highest scoring backstories that we rated as the most suitable for use by general web searchers, e.g. not requiring specific geographic knowledge or of specific events.

\begin{figure}[t]
\begin{flushleft}
    {\tt It is late, but you can't get to sleep because a sore throat has taken hold and it is hard to swallow. You have run our of cough drops, and wonder if there are any folk remedies that might help you out until morning.}
\end{flushleft}
    \caption{Example backstory selected from UQV100 test collection.}
    \label{backstory}
\end{figure}

Figure \ref{backstory} shows an example backstory selected for use in our study. We chose to use the UQV100 backstories since they relate to more recently created queries for web search tasks.

\subsection{Experimental Procedure}

Participants in our study were required to complete a search session consisting of multiple search tasks based on our selected backstories. As part of their search session they had to complete a questionnaire before, during and after undertaking each task. In this section we first give details of the questionnaire and then the experimental setup and the procedures followed.

\subsubsection{Questionnaire}

While conducting their search the user had to complete an online questionnaire in a Google form while undertaking their search activities. 
For each search task the participant completed a questionnaire divided into four sections:

\begin{itemize}

\item{{\em Basic Information Survey\/}: Participants entered their assigned user ID, age, occupation, task ID to be undertaken.}

\item{{\em Pre-Search\/}: Details of participants preexisting knowledge with respect to topic of the search task to be undertaken.}

\item{{\em In search Questions\/}: Details of the participant's search activities, including details of the query entered, documents examined, what was learned.
}

\item{{\em Post-Search\/}: Post-search feedback from the user.}

\end{itemize}

\begin{table*}[t]

    \centering
    \begin{adjustbox}{width=1\textwidth}
    \begin{tabular}{|c|c|c|}
         \hline
Questionnaire section & Questions  \\
\hline
\multirow{3}{4em}{In search questions} & What search query did you enter?\\ 
& Why did you choose these words?\\ 
& Which results look useful to you?  (please enter the document names and links ) \\ 
& How did you decide which results are useful?  \\ 
& If you opened any documents, why and what did you learn from them? \\ 
& Did this query enable you to satisfy the search task? If not, what will be your strategy to refine the query for the next iteration? \\ 
\hline
    \end{tabular}
    \end{adjustbox}
    \caption{In search questions from the Questionnaire}
    \label{Questinaire}
\end{table*}

\subsubsection{Experimental Setup}

Participants used a 
setup of 
two computers arranged with two monitors side by side on a desk in our laboratory. One monitor was used for the search session, and the other to complete the online questionnaire. Participants carried out their search tasks using a standard Google web search engine using a Google chrome browser. The search question section of the questionnaire is provided in Table \ref{Questinaire}.
A web-tracking plugin was installed in the browser to capture details of the participant's activities with the browser. In addition, all search activities were recorded using a standard screen recorder tool to enable 
post-collection review of the user activities. Approval was obtained from our university Research Ethics Committee prior to beginning the data collection.

\subsubsection{User Behaviour Categories}

In our study we are interested to analyze user behaviour with a standard user-driven search tool in terms of the interactions they make and to seek to understand the reasons for their behaviour. In doing this, we can divide search behaviour into four categories:

\begin{itemize}

\item \textbf{User type A}: The user enters one query and selects one document from the retrieved list to fulfill the information need. 

\item\textbf{User type B}: The user enter one query and opens multiple documents as they seek to fulfill their information need. 

\item\textbf{User type C}: The user performs more than one iteration of query and inspection of results in order to fulfill their information need. 

\item\textbf{User type D}: The user performs more than one search interaction, but only opens a single document. 
This behaviour may arise if the user has difficulty expressing their information need in a single query, leading to reformulations and multiple runs, or if there is little relevant information available, or the user continues to seek additional information after a successful search run has been completed.

\item Additionally there is the possibility of the case where the user issues one or more queries, but does not select any of the retrieved items, this may indicate that either the user retrieves no relevant items, cannot identify retrieved relevant items or is able to satisfy their information need from one or more of the snippets displayed in the retrieved ranked list(s). 

\end{itemize}

In our analysis we seek to identify relationships between the user's knowledge of the topic of the search, the query entered and the usefulness of the retrieved documents.

\subsubsection{Pilot Study}

A pilot study with two undergraduate students in Computer Science was conducted using two additional backstory search tasks to see how long it took them to complete the sections of the study, gain insights into the likely behaviour of participants, and to generally debug the experimental setup. 

Participants were given printed details of the instructions for their search sessions and each backstory in printed form at the begin of each task. Results from the pilot study are not included in the analysis. 
Each of the pilot search tasks took around 30 minutes to complete. Feedback from the pilot study was used to refine the specification of the questionnaire.

Based on the result of the pilot study, each participant in the main study was assigned two of the selected 12 search task backstories with the expectation that their overall session would last around one hour. Pairs of backstories for each session were selected using a Latin square procedure.

\section{Main Study: Part 1}
\label{results1}

Participants for the main study were recruited on a voluntary basis with the majority being postgraduate students studying either on MSc or PhD programmes. All participants were affiliated with our University or another public University in Ireland. A total of 17 subjects completed a search session. One participant was found not to have followed the instructions appropriately, and their contribution was excluded from analysis. Thus there were a total of 32 search tasks completed.

\subsection{Information Seeking Behaviour}

\begin{table*}
    \centering
    \begin{adjustbox}{width=1\textwidth}
    \begin{tabular}{|c|c|c|c|c|} 
         \hline 
\textbf{Variables/User Behaviour } & \textbf{No. of Interaction} & \textbf{Average time per Search} & \textbf{No. of Documents viewed}& \textbf{No. of Search task}  \\
 & & \textbf{(in minutes)} & & \\ \hline
User Type A & 1 & 22 & 1 & 2 \\ 
User Type B & 1 & 21.7 & >1 & 8 \\ 
 User Type C & >1 & 34.6& >1 &  22  \\ 
 \hline
     \end{tabular}
    \end{adjustbox}
    \caption{Types of User Behaviour}
    \label{Table1}
\end{table*}

Table \ref{Table1}
summarizes the following features for the three classes of interactive search behaviour observed in our study:

\begin{enumerate}
    \item \textbf{No. of interactions}: The number of queries used by the participant to complete the search task.
    \item \textbf{Average time per search}: The average time take by a search with this class of interaction behaviour.
    \item \textbf{No. of documents per interaction}: No of documents opened for a query.
    \item \textbf{No. of Search tasks}: The total number of search tasks which follow the particular type of user behaviour.
\end{enumerate}

\begin{table*}[h]
\centering
\begin{tabular}{|c|c|c|c|c|c|}
\hline
\textbf{Search Stage}&\textbf{ Variables} & \textbf{Type A} & \textbf{Type B} & \textbf{Type C} \\  \hline 
Search Formulation & Interesting Topic (1-5) & 4 & 4 & 3.7 \\ \cline{2-5}
& Background Knowledge (1-5) & 2.5 & 2.8 & 2.2 \\ \cline{2-5}
& Difficulty level (1-5) & 2.5 & 2.1 & 2.9 \\ 
\hline 
Content Selection & Average of number docs  & 1 & 2.5 & 4.4 \\ 
& viewed per search & & & \\ \cline{2-5}
& Why chooses these results? & \begin{tabular}[t]{@{}l@{}}-Top three \\ -Relevance\\ 
\end{tabular} & \begin{tabular}[t]{@{}l@{}}-Top three\\ -Authenticated source \\ -Relevance\\ 
\end{tabular} & \begin{tabular}[t]{@{}l@{}}-Search recommendation\\ -Learn about topic \\ 
-Relevance to \\ search topic \\ 
\end{tabular} \\ \hline 
Interaction with Content & Average Level of Satisfaction & 9.00  & 8.66  & 8.16 \\ 
 & (1-10) & & & \\ \cline{2-5}
& Total Time Taken for search & 22 minutes & 21.7 minutes & 34.6 minutes\\
\hline 
Post-search & Understanding about the topic & 6.5  & 8 & 7.6\\ \cline{2-5}
&  Expand of Knowledge After search & 4 & 4 & 4\\ \hline                                                           
\end{tabular} \\
\caption{Flowchart of characteristic of search process \cite{Bailey:2016:UTC:2911451.2914671} by the change in knowledge structure }
\label{Table2}
\end{table*}

Table \ref{Table2} shows a summary of the information collected during the study. Based on our analysis of the data gathered, we make the following observations:

\begin{enumerate}

\item \textbf{User type A}: Users showing this behaviour spent on average of 22 minutes studying a single document. This document was able to entirely satisfy their information need, and there was no opportunity to study learning behaviour from the examination of multiple documents. This behaviour was only observed for 2 of the 32 completed search tasks, in each case for a different backstory. Thus, the search behaviour is not determined by the requirements of the backstory. The existing knowledge of the topic for these searches averaged 2.5 out of 5. In this case the user exhibited sufficient existing knowledge of the subject to be able to use the provided document snippet and other metadata to 
identify a document which is able to satisfy the information need beyond their existing knowledge of the subject \citep{vakkari2016searching}. 

\item\textbf{User type B}: The number of documents opened was on average is 2.5. The average knowledge was 2.8 out of 5. The average time spent viewing each document was approximately 9 minutes, which is less than half of the time taken by type A users examining their single document.  Users thus spent an average of 21.7 minutes examining documents. Although more common than Type A behaviour, this behaviour was still not common, and we observed this behavior for only 8 search tasks.

Type B users claim a slightly greater initial knowledge in comparison to the Type A users\footnote{It should be noted that is calculated in both cases based on very small numbers of search tasks.} which may encourage them to make greater exploration of the topic by examining multiple documents.

\item \textbf{User type C}: The average time spent on each search task was 34.6 minutes, with an average of 2.6 queries and 4.4 documents opened per search task. The average prior knowledge of the search topic was 2.2 out of 5. The average time spent on viewing each document was approximately 7.8 minutes, which is nearly one third of the time time taken by type A users and less than time taken on an average by user type B.

Type C users on an average have less prior knowledge of the search topic than Type B and Type A searchers. This motivates them to carry out multiple search interactions with reformulated queries and to open multiple documents to satisfy their information need. Type C user follow all the paradigm of change in knowledge structure (restructuring, tuning and assimilation) following the search process. \cite{vakkari2016searching}. 
This behaviour was by far the mostly commonly observed in our study being used for 22 search tasks. 

\end{enumerate}

\subsection{Analysis of Search Activities}

In this section we consider our results in terms of Vakkari's model of Search as Learning. 
From the summary of the data provided by the participants in Tables \ref{Table2} 
and review of the videos of the search sessions we can make the following observations.



\begin{enumerate}

\item \textbf{Type A searchers}: For type A searches, the user has a good knowledge of the topic of the information need, and despite the complex nature of the information need is able to form a query able to retrieve a single document which completely addresses the information, or at least does so to the extent that they regard the task as completed on the basis of reading this document. In other cases a similarly knowledgeable searcher could form an equally good query, but not be able to find a single document addressing the information need. Suitable documents were observed to come from sources such as Wikipedia or a specialist website relating to the topic of the information need.
            
In terms of the searcher's knowledge structures relating to this information need, their prior knowledge of the topic means that they will already have a structured understanding of the topic, and that their cognitive activity will consist of tuning and assimilation of the information. The fact that the required information is contained in a single document makes the user's task easier, although the time taken to engage with the document to identify the necessary information means that the user is still observed to undertake a large amount of work. With respect to conversational search, the question arises of whether an agent could improve the efficiency with which the information is accessed.
            

\item \textbf{Type B searchers}: The time spent with each document is less than the single document in type A searches, which suggests that users are able to identify the relevant material relatively quickly. The overall time for task is needed for the process of accessing detailed information from across the opened documents, and to interpret it for tuning and assimilation. There prior knowledge of the topic again means that the cognitive activities will focus on tuning and assimilation, but in this case is likely to be more demanding since it is spread across consultation with multiple documents. 
In this case, this is a more demanding activity since this takes place across multiple documents. Again the follow on question to these observations, is to consider whether a conversational agent might be given the competence to improve the efficiency of the user's engagement with this information to satisfy their information need. There would appear to be scope here for a conversational agent to assist the searcher in their engagement with snippets and documents as they seek the address their information need.


\begin{figure}
\begin{flushleft}
    1st Interaction \\
    {\bf User Query 1\/}: {\tt folk remedies for sore throats\/} \\
    No of Documents viewed: 1 \\
    {\bf Doc Title\/}: {\tt 10 Natural Home Remedies for Sore Throat - Global Healing Center} \\
    2nd Interaction \\
    {\bf User Query 2\/}: {\tt folk remedies to help soothe a sore throat} \\
    No of Documents viewed: 1 \\
    {\bf Doc Title\/}: {\tt 22 Natural Sore Throat Remedies to Help Soothe the Pain} 
\end{flushleft}
    \caption{Queries issued in one search session for example backstory shown in Figure \ref{backstory}.}
    \label{queries}
\end{figure}

\item \textbf{Type C searchers}: Participants on average show less knowledge of the topic of the search task. Figure \ref{queries} shows the queries issued and summary responses in one of the search sessions for the backstory shown in Figure \ref{backstory}. To satisfy the information, these searchers need to issue more than one query and to engage with documents retrieved across more one query. These findings are consistent in that the searcher's initial query may not be sufficiently well informed to enable it to retrieve all the required information in a single run. It may also be that the multi-faceted nature of some of these search tasks means that no single query would be able to obtain all the required information.


We analyzed this point in more detail in our study. The learning activities in this case are likely to be more complex involving a combination of restructuring, tuning and assimilation as the searcher progresses through the completion of the search task. In completing the search tasks, these participants are essentially engaging in a limited dialogue with the search engine. In addition to the exploration of conversational opportunities for the search considered for Type A and Type B searches, for Type C, we can also consider how an agent can support the search through the multiple query phases.

Examination of the contents of both the initial and reformulated queries showed that most of the content or topic related words contained in them are found in the written backstories. A notable point related to the contents of the reformulated queries is that they contain very few occurrences of words found in documents opened in the previous search. On consideration allowing participants to keep the written backstory visible while carrying out the search does not represent the operational situation for many web searches where the user relies on recall of related words to form queries, and may be more reliant on words from retrieved documents in the reformulation. 

Thus, we repeated our study providing participants with the backstory, and then removing it from their view while carrying out their search. We report results and analysis of this further study in the next section.


\end{enumerate}

\section{Main Study: Part 2}
\label{results2}

\begin{table*}
    \centering
    \begin{adjustbox}{width=1\textwidth}
\begin{tabular}{|c|c|c|c|c|} 
         \hline 
\textbf{Variables/User Behaviour } & \textbf{No. of Interaction} & \textbf{Average time per Search} & \textbf{No. of Documents viewed}& \textbf{No. of Search task}  \\
 & & \textbf{(in minutes)} & & \\ \hline
User Type A & 1 & 23.2 & 1 & 5 \\ 
User Type B & 1 & 22.7 & >1 & 11 \\ 
User Type C & >1 & 33.3 & >1 &  22  \\\hline
    \end{tabular}
    \end{adjustbox}
    \caption{Types of User Behaviour without back story view}
    \label{Table1a}
\end{table*}

\begin{table*}[h]
\centering
\begin{tabular}{|c|c|c|c|c|c|}
\hline
\textbf{Search Stage}&\textbf{ Variables} & \textbf{Type A} & \textbf{Type B} & \textbf{Type C} \\  \hline 
Search Formulation & Interesting Topic (1-5) & 3.8 & 3.5 & 3.5 \\ \cline{2-5}
& Background Knowledge (1-5) & 2 & 1.9 & 2.3 \\ \cline{2-5}
& Difficulty level (1-5) & 2.6 & 2.7 & 2.6 \\ 
\hline 
Content Selection & Average of number docs  & 1 & 3.7 & 4 \\ 
& viewed per search & & & \\ \cline{2-5}
& Why chooses these results? & \begin{tabular}[t]{@{}l@{}}-Top three \\ -Relevance\\ 
\end{tabular} & \begin{tabular}[t]{@{}l@{}}-Top three\\ -Authenticated source \\ -Relevance\\ 
\end{tabular} & \begin{tabular}[t]{@{}l@{}}-Search recommendation\\ -Learn about topic \\ 
-Relevance to \\ search topic \\ 
\end{tabular} \\ \hline 
Interaction with Content & Average Level of Satisfaction & 9.00  & 8.00  & 8.18 \\ 
 & (1-10) & & & \\ \cline{2-5}
& Total Time Taken for search & 23.2 minutes & 22.72 minutes & 33.3 minutes\\
\hline 
Post-search & Understanding about the topic & 7  & 7 & 7.6\\ \cline{2-5}
&  Expand of Knowledge After search & 3.8 & 3.7 & 3.9\\ \hline                                          \end{tabular} \\
\caption{Flowchart of characteristic of search process \cite{Bailey:2016:UTC:2911451.2914671} by the change in knowledge structure without back story view}
\label{Table4}
\end{table*}

A total of 19 participants completed the second study, and completed 38 search tasks. 11 subjects participated in both studies. These participants were assigned different search tasks in the second study. Using some of the same participants in both studies allowed us to compare their behaviour across both experimental conditions.

\begin{table*}[h]
    \centering
\begin{tabular}{ |c|c|c| } 
 \hline
 \textbf{Search condition} &  \textbf{Same search behaviour in each session} &  \textbf{Different search behaviour in each session} \\\hline 
 With backstory & 8 & 8 \\ 
 Without backstory & 12 & 7 \\ 
 \hline
\end{tabular}
    \caption{Types of search behaviour with search session }
    \label{Table3}
\end{table*}


\subsection{Information Seeking Behaviour}

Results of our second study with the revised backstory condition are shown in Tables \ref{Table1a} and \ref{Table4}. Analyzing the results in terms of type A, type B and type C behaviour, we now observe the following:

\begin{enumerate}

\item \textbf{User type A}: Participants spent on average of 23.2 minutes studying a single selected document. They were observed to spend more time reading the document in the new condition without the backstory. This behaviour was only observed for 5 of the 38 completed search tasks, in each case for a different backstory. The existing knowledge of the topic for these searches averaged 2 out of 5. 

\item\textbf{User type B}:  The number of documents opened was on average is 3.7 which is higher than in the backstory condition. The average time spent viewing each document was approximately 6.1 minutes, which is less than 
the time taken by type A users examining their single document.  Participants thus spent an average of 22.7 minutes examining documents which is also longer than in the backstory condition. 
This behaviour was observed for 11 search tasks.
The background knowledge was 1.9 out of 5 which is below average, and 
again slightly less initial than 
Type A users, which may encourage them to read more documents. 


\item\textbf{User type C}: The average time spent on each search task was 33.36 minutes, with an average of 3.1 queries and 4 documents opened per search task. The average prior knowledge of the search topic was 2.3 out of 5. The average time spent on viewing each document was approximately 8.3 minutes, which is nearly one third of the time time taken by type A users in both conditions and more than the time taken on average by user type C in condition 1 and type B on condition 2. The taken by C user in condition 2 is less than time taken by User B in condition 1.Type C users have on average more prior knowledge of the search topic than Type B and Type A searchers in condition 2.

This behaviour was the most commonly observed in our study being used for 22 search tasks. In the case of Type C users, It is observed that Type C Behavior was different in condition 2.

\end{enumerate}

We summarize overall findings from our study in Tables \ref{Table2} and Table \ref{Table4}. We also observed different search behaviour in search sessions shown in Table \ref{Table3}. From these results and review of the videos of the search sessions we can make the following observations.

\begin{enumerate}

\item \textbf{Type A searchers}: We observed the difference in the parameters of Type A searchers in both conditions. Type A searchers in the second have less interest, less knowledge and greater difficulty in completing the task than Type A searchers in the backstory condition. 


\item \textbf{Type B searchers}: The time spent with each document is less than the single document in type A searches, which suggests that they are able to identify the relevant material relatively quick than B in condition 1 and Type A users in both the conditions. The type B user in condition 2 refer more documents with the less average reading time of each document. This shows type B users were reading more number of documents and spending less time in reading the document. This also reflect user might able to structure and tune his knowledge by shallow reading the multiple documents. This behaviour direct towards the conversational approach of Information retrieval.

\item \textbf{Type C searchers}: Type C users in no backstory condition spent less time  reading each document in comparison to other user behaviours in both conditions. Type C users claimed to have very good background knowledge with above average interest in the topic. 
On average type C viewed four documents per session. 

\end{enumerate}

\section{Summary Analysis}


In summary, we find that the participants appear to have to expend more effort in completing their search in the second no backstory. More interestingly, the reformulated queries for the second and subsequent queries in the Type C tasks make much more use of words appearing in retrieved documents rather than those in the written backstories. This is not really surprising since the retrieved documents provide the searcher with a ready source of topically relevant words which they can choose from to improve or revise their query statement.

With respect to conversational search, this finding is interesting because we could consider the incorporating a conversational agent in the analysis of the retrieved documents and selection of words or the automated reformulation of the query. This could go beyond recommendation of words for query expansion in current IR systems with the agent taking a more proactive role.

Ideally the agent should consider the features of the content actually engaged with by the searcher, their original query and recommended expansion words which they select, to determine the searcher's knowledge of the topic under investigation and their interests within this topic, and use this to support the search.

By taking on tasks currently driven by the user, such as examination of documents to determine how to reformulate queries, and actively personalising the behaviour of the search tool, taking account of the searchers subject knowledge and interests, a conversational agent should be able to reduce overall load on the user in completing their search task. Information seeking models can provide a basis for a structure for the behaviour of such agents to support search, taking account of the behaviour of the searcher as they learn about the subject under consideration while they work to address their information need.

\section{Conclusions and Directions for Further Work}
\label{conclusions}

In this paper we studied user search with a standard commercial search engine for cognitively demanding search tasks, with the objective of informing the development of conversational search systems for use in these tasks. We performed the experiment in two conditions, with back story and without back story. We categorized interactive information seeking behaviour into four types. We observed the first three types in our study. Analysis and observation of the user's behaviour for these three types can broadly be explained in terms of Vakkari's learning model of search in which searchers develop, refine and assimilate knowledge gained within the search.

As the next stage of our work, we plan to repeat this study with a refined experimental procedure to better simulate user's working with a personal information need. We note that the participants in the study were advanced level students, most of them with a technical background in Computer Science. For a further study we plan to recruit participants with more diverse backgrounds to examine their search behaviour.  We also plan to explore the extension of the study design to incorporate elements of conversational support into the search process to gauge searcher response and gain feedback. We will also examine how conversational agents for deployment in tasks of this type might be implemented.


\section{Acknowledgements}

This work was supported by Science Foundation Ireland as part of the ADAPT Centre at DCU (Grant No. 13/RC/2106) (\url{www.adaptcentre.ie}).

\bibliographystyle{abbrv}
\bibliography{reference}
\appendix

\end{document}